\documentclass[11pt]{article}
\usepackage{amssymb,amsmath}
\usepackage{graphicx}
\usepackage{epsfig}
\usepackage{epstopdf}
\usepackage{latexsym}
\usepackage{psfrag}
\usepackage{booktabs}
\usepackage{bbm}
\usepackage[toc]{appendix}
\usepackage{color}
\usepackage{datetime}
\usepackage[
      colorlinks=true,
      linkcolor=darkblue,  
      urlcolor=blue,    
      filecolor=blue,     
      citecolor=red,
      linktocpage=true,
      pdfstartview=FitV,
      bookmarksopen=true    
      ]{hyperref}

\definecolor{darkblue}{rgb}{0.2, 0, 0.8}

\numberwithin{equation}{section}

\usepackage{dsfont}
\usepackage{amsfonts}
\usepackage{amssymb}
\usepackage{amsmath}
\usepackage{graphicx}
\usepackage{slashed}
\usepackage{setspace}

\usepackage{color}

\newcommand{\reef}[1]{(\ref{#1})}

\newcommand{\be}{\begin{equation}}
\newcommand{\ee}{\end{equation}}

\def\be{\begin{equation}}
\def\ee{\end{equation}}
\def\bea{\begin{eqnarray}}
\def\eea{\end{eqnarray}}
\def\ba{\begin{array}}
\def\ea{\end{array}}
\def\bd{\begin{displaymath}}
\def\ed{\end{displaymath}}

\def\>{\rangle} 
\def\<{\langle} 
\def\Dsl{D \hskip-.6em \raise1pt\hbox{$ / $ } }
\def\to{\rightarrow}

\newcommand{\eps}{\epsilon}

\begin{document}  


\begin{titlepage}

 \begin{flushright}
{\tt MCTP-15-12} 
\end{flushright}

\vspace{2.3cm}

\begin{center}
{\Large \bf Exact results for corner contributions to the} \\[2.5mm]
{\Large \bf  entanglement entropy and R\'enyi entropies  }\\[2.5mm]
{\Large \bf of free bosons and fermions in 3d} \\

\vspace*{1.2cm}

{\bf Henriette Elvang and Marios Hadjiantonis}
\medskip

Randall Laboratory of Physics, Department of Physics,\\
University of Michigan, Ann Arbor, MI 48109, USA
\bigskip

\bigskip
\tt{elvang@umich.edu and mhadjian@umich.edu}  \\
\end{center}

\vspace*{0.1cm}

\begin{abstract}  
In the presence of a sharp corner in the boundary of the entanglement region, the entanglement entropy (EE) and R\'enyi entropies for 3d CFTs have a logarithmic term whose coefficient, the corner function, is scheme-independent. In the limit where the corner becomes smooth, the corner function vanishes quadratically with coefficient $\sigma$ for the EE  and $\sigma_n$ for the R\'enyi entropies. For a free real scalar and a free Dirac fermion, we evaluate analytically the integral expressions of Casini, Huerta, and Leitao to derive exact results for $\sigma$ and $\sigma_n$ for all $n=2,3,\dots$. The results for $\sigma$ agree with a recent universality conjecture of Bueno, Myers, and Witczak-Krempa that $\sigma/C_T = \pi^2/24$ 
in all 3d CFTs, where $C_T$ is the central charge. For the R\'enyi entropies, the ratios $\sigma_n/C_T$ do not indicate similar universality. 
However, in the limit $n \to \infty$, the asymptotic values satisfy a simple relationship and equal  $1/(4\pi^2)$  times the asymptotic values of the free energy of free scalars/fermions on the $n$-covered 3-sphere.

\end{abstract}

\end{titlepage}

\section{Introduction and Results}
\label{sec:Introduction}

For a 3d conformal field theory (CFT) in the ground state, the entanglement entropy $S$ for a region whose boundary has a sharp corner with angle $\theta$ can be written as 
\be
  S= B \,\frac{L}{\eps} - a(\theta) \ln \Big( \frac{L}{\eps}\Big) + O(1)\,.
\ee
Here $L$ is a length scale associated with the size of the entangling region, $\eps$ is a short distance cutoff, and $B$ is a non-universal constant. The {\em corner contribution} to the entanglement entropy is the scheme-independent positive function $a(\theta)$ of the opening angle $\theta$ \cite{Fradkin:2006mb,Casini:2006hu,Hirata:2006jx}. Since the entanglement entropy of the region equals that of the complement region, the corner contribution satisfies $a(2\pi - \theta) = a(\theta)$. 
If the curve bounding the entangling region is smooth, the logarithmic term is absent, hence $a(\theta)$ must vanish in the limit $\theta \to \pi$ and it does so quadratically as
\be
 \label{defsigma}
  a(\theta) = \sigma \, (\theta - \pi )^2 + \dots 
  ~~~~\text{for}~~~~
  \theta \to \pi\,.
\ee
The value of the {\em corner coefficient} $\sigma$ depends on the theory. 

For the theory of a free real scalar or a Dirac fermion, Casini, Huerta, and Leitao \cite{Casini:2008as,Casini:2009sr,Casini:2006hu} derived expressions that give 
$a(\theta)$  implicitly in terms of some rather involved integrals. In the limit, $\theta \to \pi$ one can extract double-integral expressions for the corner coefficient $\sigma$  in \reef{defsigma}. These integrals have been evaluated  numerically \cite{Casini:2009sr,Bueno:2015rda} and the results indicate  that the exact values are \cite{Bueno:2015rda}
\be
\boxed{
  \label{sigmaBFval}
  \sigma^{(B)} = \frac{1}{256} 
  ~~~~\text{and}~~~~
  \sigma^{(F)} = \frac{1}{128}\,}
\ee
for the free boson and free fermion, respectively. 
  
Bueno, Myers, and Witczak-Krempa  \cite{Bueno:2015rda} conjectured that the ratio of the coefficient $\sigma$ in \reef{defsigma} to the central charge $C_T$ is universal in 3d CFTs and that it takes the value
\be
  \label{conj}
  \hspace{-1cm}
  \text{conjecture  \cite{Bueno:2015rda}:}
  ~~~~~~~~\frac{\sigma}{C_T} = \frac{\pi^2}{24} \,.
\ee
 The conjecture \reef{conj} has passed non-trivial holographic tests for gravity models with a family of higher derivative corrections \cite{Bueno:2015rda,Bueno:2015xda}. The central charge $C_T$ is defined as the coefficient of the vacuum 2-point function of the stress tensor (see eq.~(3) in \cite{Bueno:2015rda}). For free bosons and fermions, Osborn and Petkou \cite{Osborn:1993cr}  found that $C_{T}^{(B)} =3/(32\pi^2)$ and $C_{T}^{(F)} =3/(16\pi^2)$  in 3d. So with the values \reef{sigmaBFval}, the ratio $\sigma/C_T$ is indeed $\pi^2/24$ for both free bosons and fermions.  
 
In this paper, we evaluate analytically the integral expressions \cite{Bueno:2015rda} of Casini, Huerta, and Leitao \cite{Casini:2008as,Casini:2009sr,Casini:2006hu} for $\sigma^{(B)}$ and $\sigma^{(F)}$   
and prove that their exact values are indeed those in \reef{sigmaBFval}. This verifies the universality conjecture \reef{conj} for the case of free bosons and fermions. One way of viewing the conjecture is simply as the statement that the corner coefficient $\sigma$ in \reef{defsigma} does not contain independent information about the CFT, but is fixed in terms of the central charge $C_T$.

Turning to the R\'enyi entropies $S_n$, one can define a similar corner contribution $a_n(\theta)$ which in the smooth limit $ \theta \to \pi$ goes to zero as $a_n(\theta) = \sigma_n \, (\theta - \pi )^2 + \dots$ for $n=2,3,4,\dots$. (The $n \to 1$ limit of the R\'enyi entropy  is the entanglement entropy.)   
It is not known if $\sigma_n/C_T$ has any universal properties. 

We calculate $\sigma_n$ analytically for the free boson and free fermion using integral expressions for $\sigma_n$ derived in \cite{Casini:2008as,Casini:2009sr,Casini:2006hu}.\footnote{We are grateful to Horacio Casini for sharing with us the integral expression for $\sigma_n^{(F)}$.} For the free scalar we find 
\be 
  \label{sigman-res}
  \boxed{
  \sigma_n^{(B)} =\sum_{k=1}^{n-1} \frac{k(n-k)(n-2k) \tan\big (\frac{k\pi}{n}\big)}{24 \pi \,n^3 (n-1)} \,. 
  }
\ee
Note that when $n$ is even, the contribution from $k = n/2$ must be taken carefully using $\lim_{k \to n/2} (n-2k) \tan\big (\frac{k\pi}{n}\big) = 2n/\pi$. 

The result for the free fermion is
\be 
  \label{sigman-resF}
  \boxed{
  \sigma_n^{(F)} = \sum_{k=-(n-1)/2}^{(n-1)/2} \frac{k(n^2-4k^2) \tan\big (\frac{k\pi}{n}\big)}{24 \pi \,n^3 (n-1)} \,, 
  }
\ee
where sum is to be taken in integer steps from $-\tfrac{n-1}{2}$ to $\tfrac{n-1}{2}$.

For low values of $n$, the finite sums of the trigonometric functions in \reef{sigman-res}  and \reef{sigman-resF}  simplify quite nicely. The results for first nine values of $\sigma_n$ are
\be
  \nonumber
  \begin{array}{llllccccc}
      n ~~& \sigma_n^{(B)}
      && \sigma_n^{(F)}
   \\[2mm] 
     2 
     & \frac{1}{48 \pi^2} 
      { \scriptstyle ~\approx~ 0.00211086}
     &&
     \frac{1}{64 \pi} 
      {\scriptstyle ~\approx~0.00497359}
     \\[2mm]
     3 
     & \frac{1}{108 \sqrt{3} \pi}  
        {\scriptstyle ~\approx~0.00170163}
     &&     
     \frac{5}{216 \sqrt{3} \pi}   
        {\scriptstyle ~\approx~0.00425408}     
     \\[2mm]
     4&
     \frac{8+3 \pi}{1152 \pi^2} 
       {\scriptstyle ~\approx~ 0.00153255}
     &&
     \frac{1+6 \sqrt{2}}{768 \pi} 
       {\scriptstyle ~\approx~0.00393133}
     \\[2mm]
     5& 
     \frac{\sqrt{25-2 \sqrt{5}}}{1000 \pi} 
       {\scriptstyle ~\approx~0.00144219}
     &&
     \frac{\sqrt{425+58 \sqrt{5}}}{2000 \pi} 
       {\scriptstyle ~\approx~0.00374840}
     \\[2mm]
     6
     &\frac{81+34 \sqrt{3} \pi}{19440 \pi^2} 
      {\scriptstyle ~\approx~0.00138643}
     &&
     \frac{261+20 \sqrt{3}}{25920 \pi} 
      {\scriptstyle ~\approx~0.00363061}
     \\[2mm]
     7
     &\frac{2 \cot (\frac{\pi}{14})+ 5 \cot (\frac{3\pi}{14})+5 \tan (\frac{\pi}{7})}{4116\pi} 
      {\scriptstyle ~\approx~0.00134874} 
     && 
     \frac{13 \cot (\frac{\pi}{14})+ 22 \cot (\frac{3\pi}{14})+15 \tan(\frac{\pi}{7})}{8232\pi} 
      {\scriptstyle ~\approx~0.00354841} 
     \\[2mm]
     8
     &\frac{32+ 9\pi (1+ 2\sqrt{2}) }{10752 \pi^2} 
       {\scriptstyle ~\approx~0.00132161}
     && 
     \frac{1+6 \sqrt{2}+ 4 \sqrt{274+ 17\sqrt{2}} }{7168 \pi} 
       {\scriptstyle ~\approx~0.00348777}
     \\[2mm]
     9
     &\frac{27 \sqrt{3} + 10\cot (\frac{\pi}{18})+ 28 \tan (\frac{\pi}{9})+35 \tan(\frac{2\pi}{9})}{34992\pi} 
      {\scriptstyle ~\approx~0.00130116}
    && 
    \frac{135 \sqrt{3} + 68\cot (\frac{\pi}{18})+ 77 \tan (\frac{\pi}{9})+130 \tan(\frac{2\pi}{9})}{69984\pi} 
      {\scriptstyle ~\approx~0.00344118}
    \\[2mm]
    10
    & \frac{125+6 \pi\sqrt{ 565 + 142 \sqrt{5}} }{54000 \pi^2}
      {\scriptstyle ~\approx~0.00128522}
        && 
    \frac{5+ 300 \sqrt{5}+4 \sqrt{ 425 + 58 \sqrt{5}} }{72000 \pi}
      {\scriptstyle~\approx~ 0.00340427}
    && \\[2mm]
  \end{array}
\ee   
In the case of the scalar, the exact $n=2,3$ results were guessed by the authors of \cite{Bueno:2015rda} based on their high precision numerical evaluation of the integrals. 

Since the ratios of the central charges of free fermions and bosons differ only by a factor of 2,  universality of the ratio $\sigma_n/C_T$ would require that  $\sigma_n^{(B)}/\sigma_n^{(F)}$ obeys some simple, possibly $n$-dependent, relation. Based on our results above, there is no hint of such a simple relationship. Of course to fully exclude this, one would need values of $\sigma_n$ for other 3d CFTs. 

\begin{figure}[t]
\centerline{
\includegraphics[width=7cm]{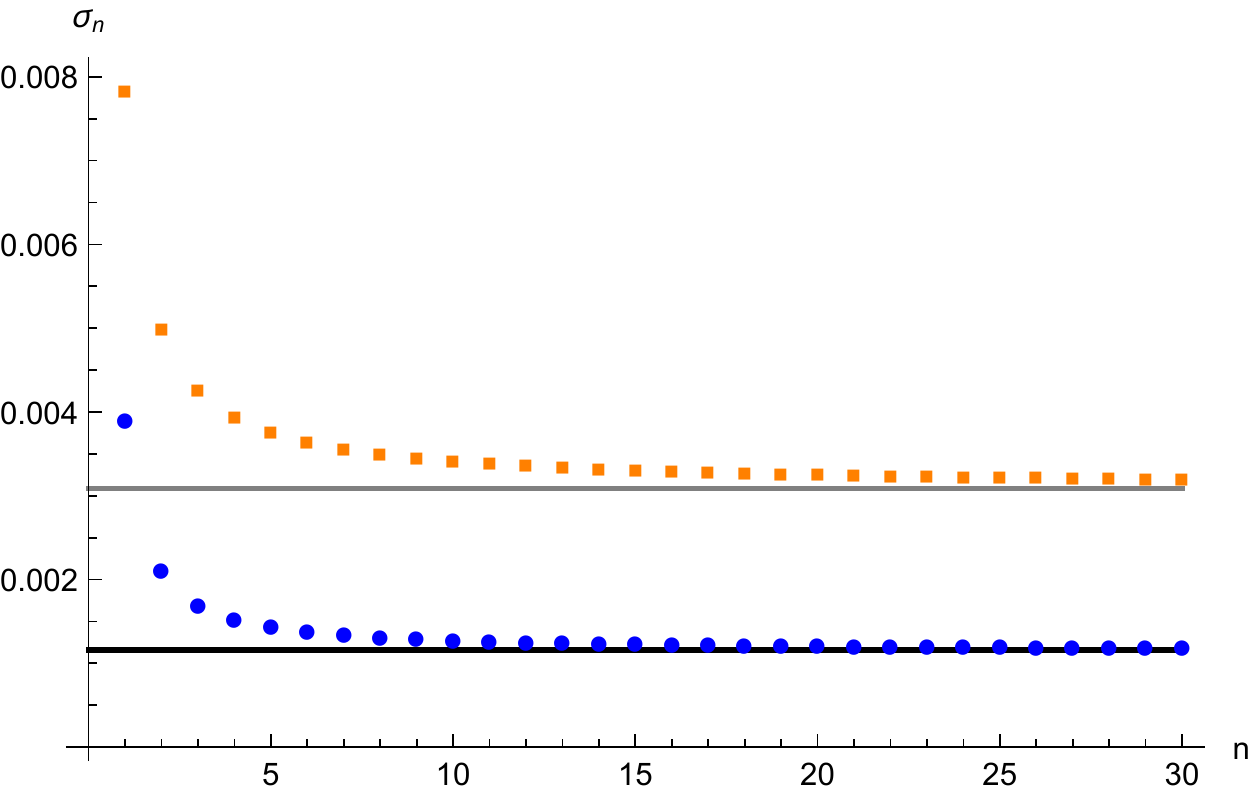}
~~~~~
\includegraphics[width=7cm]{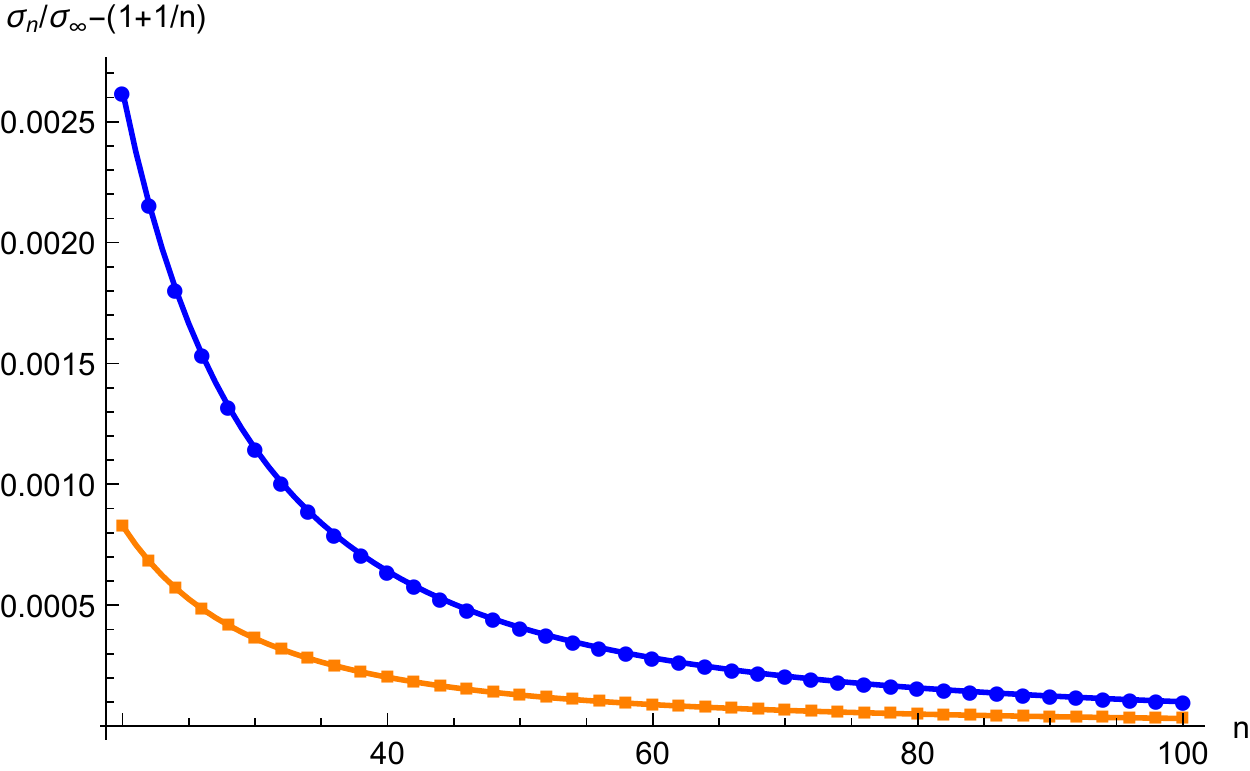}
}
\caption{\small {\em Left:} Plot showing that $\sigma_n$ decreases monotonically from  the entanglement entropy value included for $n=1$ to the asymptotic value $\sigma_\infty$ for free scalars (blue) and free Dirac fermions (maize squares). 
The asymptotic values $\sigma_\infty^{(B)} = \tfrac{3 \zeta(3)}{32 \pi^4}\approx 0.0011569$ (black) and $\sigma_\infty^{(F)} = \tfrac{\zeta(3)}{4 \pi^4}\approx 0.00308507$ (gray) are indicated as horizontal lines.  {\em Right:} The plot illustrates our numerical fit $\sigma_n =\sigma_\infty \big ( 1+ \frac{b_1}{n} + \frac{b_2}{n^2} + \frac{b_3}{n^3} + \dots \big)$, for which we find $b_1=b_2=b_3=1$ for the free scalar, and $b_1=1$ and $b_2 = b_3 =1 - \frac{\pi^2}{12\zeta(3)} \approx  0.31578$ for the free fermion; the solid curves are $\frac{b_2}{n^2} + \frac{b_3}{n^3}$ for those respective values of $b_2$ and $b_3$.}
\label{sigmaplot}
\end{figure}

As a function of $n$, the R\'enyi corner coefficient $\sigma_n$ decreases monotonically, as shown on the left in figure \ref{sigmaplot}.  When $n$ is large, $\sigma_n$ asymptotes to a constant value,  which we calculate analytically:
\be
  \boxed{
  \sigma_\infty^{(B)} = \frac{3 \zeta(3)}{32 \pi^4} 
  ~~~~\text{and}~~~~
  \sigma_\infty^{(F)} = \frac{\zeta(3)}{4 \pi^4} 
  \,.
  }
\ee
The appearance of the Riemann zeta-function is intriguing since $\zeta(3)\,\approx 1.20206$ 
also shows up in the free energies and R\'enyi entropies for free scalars/fermions on a 3-sphere, as shown by Klebanov, Pufu, Sachdev, and Safdi \cite{Klebanov:2011uf}.  Specifically, the free energy of a free real scalar or free fermion on an $n$-covered 3-sphere behaves as\footnote{The authors of \cite{Klebanov:2011uf} work with a complex scalar, so the free energy there is twice that of a real scalar.} 
$\mathcal{F}_n \to n \mathcal{F}_\infty$ for $n\to \infty$ with   
\be  
  \mathcal{F}_\infty^{(B)} =\frac{3 \zeta(3)}{8 \pi^2}  
  ~~~~\text{and}~~~~
  \mathcal{F}_\infty^{(F)} =\frac{\zeta(3)}{\pi^2} \,.
\ee
Thus, for both free scalars and fermions  we have
\be
  \label{sigmaF}
  \sigma_\infty^{(B/F)} = \frac{1}{4\pi^2}
  \mathcal{F}_\infty^{(B/F)} \,.
\ee
For finite $n$, there is no apparent relation between $\mathcal{F}_n$ and $\sigma_n$, however there are some similarities in the subleading large-$n$ behaviors, as we discuss in section \ref{sec:AB}. 
The plot on the right in figure \ref{sigmaplot} shows the large-$n$ behaviors of the R\'enyi corner coefficients $\sigma_n$. 
A priori it is not clear if there is any relation at large $n$ between $\sigma_n$ and $\mathcal{F}_n$, but it would be curious to test \reef{sigmaF} in other examples. 

The remainder of the paper details the derivations of the results summarized above. In section \ref{sec:EE}, we derive the results \reef{sigmaBFval} for the entanglement entropy corner coefficient $\sigma$. We then evaluate the R\'enyi entropy corner coefficients $\sigma_n$ in section \ref{sec:RE}. In section \ref{sec:AB}, we discuss the asymptotic behavior at large $n$.

\section{Evaluation of the EE integrals}
\label{sec:EE}

In this section we describe the procedure for analytically evaluating the integrals  for the coefficients $\sigma^{(B)}$ and $\sigma^{(F)}$ of the entanglement entropy.
Our starting point is the integrals \cite{Casini:2008as,Casini:2009sr,Casini:2006hu} presented 
in equations (B1)-(B3) of \cite{Bueno:2015rda}. 
After a change of integration variable from $m$ to $\mu = \sqrt{4 m^2 - 1}$, the integrals take the form
\be
\begin{split}
\sigma^{(B)} &=~ -\frac{1}{2} \int_0^\infty d\mu \int_0^\infty db\, \mu^2 H a \left( 1 - a \right) 
\frac{\pi}{\cosh^2 \big( \pi b \big)} \,, \\
\sigma^{(F)} &=~ - \int_0^\infty d\mu \int_0^{\infty} db \left[ \mu^2 H a \left( 1 - a \right) - \frac{\mu F}{4 \pi} \right] \frac{\pi}{\sinh^2 \big( \pi b \big)}\,,
\end{split}
\ee
where $a = 1/2 + i b$ for the scalar  and $a = i b$ for the fermion. The functions $H$ and $F$ are defined as
\be
\label{Hdef}
H = - \frac{T}{2} \left( \frac{c}{h} X_1 + \frac{1}{c} X_2 \right) + \frac{1}{16 \pi a \left( a - 1 \right)}\,,
~~~~~~~~~~
F = - \frac{F_1}{F_2} \,,
\ee
with\footnote{We simplified the expression for $F_1$ in \cite{Bueno:2015rda} by writing it in terms of $H$.}
\be
\begin{split}
F_1 &=~  4 \pi c h H a \left( 1-a  \right) \left[ \left( 2 a - 1 \right)^2 + \mu^2  \right] - \frac{1}{4} c h^2 \left( 
\mu^2 + 1 \right) \,,
\label{F1}
\\
F_2 &=~ \frac{c h \left[ \left( 2 a - 1 \right)^2 + \mu^2 \right]^2}{2 \left( 2 a - 1 \right) \mu} \,.
\end{split}
\ee
The functions $h$, $c$, $X_1$, $X_2$, and $T$ are defined as follows:
\bea
\nonumber
h &=& \frac{2 \left( \mu^2 + (2a-1)^2 \right) \sin^2 \left( \pi a \right)}{\left( \mu^2 + 1 \right) \left( \cos \left( 2 \pi a \right) + \cosh \left( \pi \mu \right) \right)}\,,
\\[1mm]
\nonumber
c &=& \frac{2^{2 a} \pi a \left( 1 - a \right) \sec \left( \pi a + \frac{i \pi \mu}{2}  \right) \Gamma\! \left( \frac{3}{2} - a + \frac{i \mu}{2} \right)}{\sqrt{\mu^2 + 1} \left( \Gamma\! \left( 2 - a \right) \right)^2 \Gamma\! \left( -\frac{1}{2} + a + \frac{i \mu}{2} \right)}\,,
\\[1mm]
\nonumber
X_1 &=& - \frac{\Gamma\! \left( - a \right) \left[ \pi \sinh \left( \frac{\pi \mu}{2} \right) + i \cosh \left( \frac{\pi \mu}{2} \right) \left( \psi\! \left( \frac{1}{2} + a + \frac{i \mu}{2} \right) - \psi\! \left( \frac{1}{2} + a - \frac{i \mu}{2} \right) \right) \right]}{2^{2 a + 1} \mu \Gamma\! \left( a + 1 \right) \Gamma\! \left( \frac{1}{2} - a + \frac{i \mu}{2} \right) \Gamma\! \left( \frac{1}{2} - a - \frac{i \mu}{2} \right) \big( \cos \left( 2 \pi a \right) + \cosh \left( \pi \mu \right) \big)}\,,
\\[1mm]
\nonumber
X_2 &=& \big(X_1 \text{ with } a \text{ replaced by } (1-a) \big)\,, \\[1mm]
\label{hcx1x2T}
T &=& \frac{1}{2} \sqrt{h \left[ \left( h + 1 \right) \left( \mu^2 + 1 \right) - 4 a \left( 1 - a \right) \right]}\,.
\eea
Here $\psi$ denotes the digamma function, $\psi(z) = \tfrac{d}{dz} \ln \Gamma(z)$.

Our first line of attack involves calculating the quantities $c X_1 / h$ and $X_2 / c$ that appear in $H$ in \reef{Hdef}. Beyond  the immediate cancellations that occur in these ratios, 
one can perform further simplifications using identities involving gamma functions.
Namely, one can use the recurrence relation
\be
\Gamma\! \left( 1 + z \right) = z \Gamma\! \left( z \right)
\label{gamma-recurrence}
\ee
and the reflection relation
\be
\Gamma\! \left( 1- z \right) \Gamma\! \left( z \right) = \frac{\pi}{\sin \left( \pi z \right)}.
\label{gamma-reflection}
\ee
Surprisingly, all the gamma functions cancel after a series of such substitutions, giving
\be
\begin{split}
\frac{c}{h} X_1 &= \frac{\sqrt{\mu^2 + 1} \csc \left( \pi  a \right)}{16 \pi \mu \left( a - 1 \right) a} 
\left[ \pi  \sinh \left( \tfrac{\pi \mu}{2} \right)+ i \cosh \left( \tfrac{\pi \mu}{2} \right) \left( \psi\! \left( \tfrac{1}{2} + a + \tfrac{i \mu}{2} \right)- \psi\! \left( \tfrac{1}{2} + a - \tfrac{i \mu}{2} \right)\right)\right],
\\[2mm]
\frac{1}{c} X_2 &= \frac{\sqrt{\mu^2 + 1} \csc \left( \pi  a \right)}{16 \pi \mu \left( a - 1 \right) a} 
\left[ \pi  \sinh \left( \tfrac{\pi \mu}{2} \right)+ i \cosh \left( \tfrac{\pi \mu}{2} \right) \left( \psi\! \left( \tfrac{3}{2} - a + \tfrac{i \mu}{2} \right)- \psi\! \left( \tfrac{3}{2} - a - \tfrac{i \mu}{2} \right)\right) \right] .
\end{split}
\ee
It is suggestive that the pre-factors and the form of these two results are the same.
We then proceed by adding them together as  in \reef{Hdef}.
The linear combination of digamma functions that appears in the result can be simplified using properties easily derived from (\ref{gamma-recurrence}) and (\ref{gamma-reflection}).
In the form that is useful for our purpose, these identities are
\[
\psi\! \left( \tfrac{3}{2} - a \pm \tfrac{i \mu}{2} \right) = \psi\! \left( \tfrac{1}{2} - a \pm \tfrac{i \mu}{2} \right) + \frac{1}{\tfrac{1}{2} - a \pm \tfrac{i \mu}{2}}
\] 
and
\[
\psi\! \left( \tfrac{1}{2} + a \pm \tfrac{i \mu}{2} \right) - \psi\! \left( \tfrac{1}{2} - a \mp \tfrac{i \mu}{2} \right) = \pi \tan \left( \pi a \pm i \tfrac{\pi \mu}{2} \right)\,.
\]  
Then the combination of $X_1$ and $X_2$ that appears in $H$ simplifies to
\be
\label{x1x2combo}
\frac{c}{h} X_1 + \frac{1}{c} X_2 = \frac{\sqrt{\mu ^2+1}}{4 \pi a (1 - a)} \left(\frac{\pi \sin (\pi  a) \sinh
   \left(\frac{\pi  \mu }{2}\right)}{\mu \left[\cos (2 \pi  a)+\cosh (\pi  \mu )\right]}-\frac{\csc (\pi  a) \cosh \left(\frac{\pi  \mu }{2}\right)}{(1-2 a)^2+\mu ^2}\right)\, .
\ee
The last ingredient we need to construct $H$ in \reef{Hdef} is $T$. Using  \reef{hcx1x2T}, it is  
\be
T = \sqrt{\frac{\big((1-2 a)^2+\mu ^2\big)^2 \sin ^2(\pi  a) \cosh ^2\left(\frac{\pi  \mu }{2}\right)}{(\mu ^2+1) \big(\cos (2 \pi  a)+\cosh (\pi \mu )\big)^2}} \,.
\ee
Further simplifications of $H$ depend on the nature of variable $a$, as we 
will see when we specialize to the cases of the free scalar and the free fermion.

\vspace{2mm}
\noindent {\bf Free scalar.} 
To proceed with the evaluation of the integral $\sigma^{(B)}$, we set $a = 1/2 + i b$ as prescribed for the free scalar. It is furthermore convenient to change integration variable $b \to b/2$. Using that both $\mu$ and $b$ are positive, the integrand of $\sigma^{(B)}$ simplifies dramatically and becomes 
\be
\sigma^{(B)} = \int_0^\infty d\mu \int_0^\infty db\ \frac{\mu \left[ \pi \left( \mu^2 - b^2 \right) \sinh (\pi \mu ) + 2 \mu \cosh (\pi  b) - 2 \mu \cosh (\pi  \mu ) \right]}{64 [\cosh (\pi b) - \cosh (\pi \mu)]^2}\,.
\ee
Next, we integrate by parts. Writing 
\be
\sigma^{(B)} = \frac{1}{64} \int_0^\infty d\mu \int_0^\infty db\ \left[ \frac{\partial}{\partial \mu} \left( \frac{\mu \left( \mu^2 - b^2 \right)}{\cosh \left( \pi b \right) - \cosh \left( \pi \mu \right)} \right) + \frac{b^2-\mu^2 }{\cosh \left( \pi b \right) - \cosh \left( \pi \mu \right)} \right] ,
\ee
we see that the boundary term vanishes and we get 
\be
\sigma^{(B)} = \frac{1}{256} \int_{-\infty}^{+\infty} d\mu \int_{-\infty}^{+\infty} db\ \frac{b^2 - \mu^2}{\cosh \left( \pi b \right) - \cosh \left( \pi \mu \right)}\,.
\ee
We have extended the limits of integration to facilitate the change of integration variables
\be
 \label{bmuxy}
 \mu = x - y
 ~~~~\text{and}~~~
 b = x + y \,.
\ee
This separates the two integrations and reduces the expression to
\be
\sigma^{(B)} = \left( \frac{1}{8} \int_{-\infty}^{+\infty} dx\ \frac{x}{\sinh \left( \pi x \right)} \right)^2 = \frac{1}{256}\,.
\ee
This completes the derivation of the result \reef{sigmaBFval} for the free scalar. 


\vspace{2mm}
\noindent {\bf Free fermion.} 
With $F_1$ given in terms of $H$ as in \reef{F1}, we have already 
 done most of the leg-work needed to compute $\sigma^{(F)}$. For the free fermion, we have to take $a = i b$ and it is again convenient to change integration variable $b \to b/2$. After putting everything together, we have
\be
\sigma^{(F)}=-\frac{1}{32} \int_0^\infty d\mu \int_0^\infty db\ \frac{\mu \big[ \pi \left( \mu^2 - b^2 - 1 \right) \sinh (\pi \mu ) - 2 \mu \cosh (\pi b) - 2 \mu  \cosh (\pi \mu ) \big]}{\left[ \cosh (\pi b) + \cosh (\pi \mu) \right]^2} \,.
\ee
We can  express the integrand as a total derivative plus remaining terms as
\be
\sigma^{(F)} = \frac{1}{32} \int_0^\infty d\mu \int_0^\infty db\ \left[ \frac{\partial}{\partial \mu} \left( \frac{\mu \left( \mu^2 - b^2 - 1 \right)}{\cosh (\pi b) + \cosh (\pi \mu)} \right) + \frac{1-\mu^2 +b^2}{\cosh (\pi b) + \cosh (\pi \mu)} \right] .
\ee
As before, the boundary term vanishes and we are left with the expression (after extending the limits of integration)
\be
\sigma^{(F)} = \frac{1}{128} \int_{-\infty}^{+\infty} d\mu 
\,db\ \frac{1-\mu^2 +b^2 }{\cosh (\pi b) + \cosh (\pi \mu)}
=\frac{1}{128} \int_{-\infty}^{+\infty} dx\, dy\, \frac{1+4 x y}{\cosh (\pi x )\cosh (\pi y )} \,.
\ee
In the last step, we changed integration variables using \reef{bmuxy}. Since $x/\cosh(\pi x)$ is odd, that part of the integral vanishes and the result is therefore simply 
\be
\sigma^{(F)} = \frac{1}{128} \left( \int_{-\infty}^{+\infty} dx\ \frac{1}{\cosh (\pi x )} \right)^2 
= \frac{1}{128}\,.
\ee
Thus we have derived the result \reef{sigmaBFval} for the free fermion.

\section{R\'enyi entropies}
\label{sec:RE}

We now proceed to calculate the corner coefficients $\sigma_n$ for the R\'{e}nyi entropies.

\vspace{2mm}
\noindent {\bf Free scalar.} 
For the scalar field,  the R\'{e}nyi corner coefficient is given by the integral (B7) in \cite{Bueno:2015rda}. 
We change of the integration variable $m$ to $\mu = \sqrt{4 m^2 - 1}$ to write it as
\be
\label{sigmanBdef}
\sigma_n^{(B)} = - \sum_{k = 1}^{n - 1} \frac{k \left( n - k \right)}{2 n^2 \left( n - 1 \right)} \int_0^\infty d\mu\ \mu^2 H_{k/n}\,,
\ee
where $H_{k/n}$ is $H$ in \reef{Hdef} with $a$ replaced by $k/n$.
With the simplified expression for $H$ from section \ref{sec:EE}, we get
\begin{multline}
\phantom{!}\!\!\!\!\!\!
\sigma_n^{(B)} = \sum_{k = 1}^{n - 1} \frac{\sin ^2\left(\frac{\pi  k}{n}\right)}{32 \pi n^2 (n-1)} 
\int_0^\infty d\mu\ \frac{\mu \left[ (n - 2 k)^2 + \mu^2 n^2 \right] \pi \sinh (\pi \mu ) - 2 \mu^2 n^2 \left[ \cos \left(\frac{2 \pi k}{n} \right) + \cosh (\pi  \mu ) \right]}{\left[ \cos \left( \frac{2 \pi k}{n} \right) + \cosh (\pi \mu ) \right]^2} \,.
\end{multline}
As before, we write the integrand as a total derivative plus the remaining terms:
\begin{multline}
\phantom{!}\!\!\!\!\!\!
\sigma_n^{(B)} = - \sum_{k = 1}^{n - 1} \frac{\sin ^2\left(\frac{\pi  k}{n}\right)}{32 \pi n^2 (n-1)} 
\int_0^\infty d\mu \left[ \frac{\partial}{\partial \mu} \left( \frac{\mu \left[ (n - 2 k)^2 + \mu^2 n^2 \right]}{\cos \left( \frac{2 \pi k}{n} \right) + \cosh (\pi \mu )} \right) - \frac{(n - 2 k)^2 + \mu^2 n^2}{\cos \left( \frac{2 \pi k}{n} \right) + \cosh (\pi \mu )} \right] \!.
\end{multline}
The boundary term vanishes and the expression simplifies to
\be
\label{sigmaBnow}
\sigma_n^{(B)} = \sum_{k = 1}^{n - 1} \frac{\sin ^2\left(\frac{\pi  k}{n}\right)}{32 \pi n^2 (n-1)} \int_0^\infty d\mu\ \frac{(n - 2 k)^2 + \mu^2 n^2}{\cos \left( \frac{2 \pi k}{n} \right) + \cosh (\pi \mu )}\,.
\ee
The contribution of $k = n/2$ is easy to calculate and is equal to 
\be
\label{speck}
\frac{1}{64 \pi \left( n - 1 \right)} \int_0^\infty d\mu\ \frac{\mu^2}{\sinh^2 \left( \frac{\pi \mu}{2} \right)} 
~=~ \frac{1}{48 \pi^2 \left( n - 1 \right)}\,.
\ee
For $k \neq n / 2$, there are contributions from two integrals:
\be
  \label{I1}
  I^{(1)}_{n;k}=\int_0^\infty \frac{d\mu}{\cos \left( \frac{2 \pi k}{n} \right) + \cosh (\pi \mu )}
  = \frac{2\tan^{-1} \left(\tan \left( \frac{\pi k}{n} \right) \right)}{\pi \sin\!\big(\frac{2k\pi}{n}\big)}
  =\frac{2}{\sin\!\big(\frac{2k\pi}{n}\big)} \times\left\{ 
  \begin{array}{ll}
\frac{k}{n}\,, & k < n / 2 \\[1mm]
\frac{k}{n} - 1\,, & k > n / 2
\end{array} \right. 
\ee
and
\bea
  \label{I2}
  I^{(2)}_{n;k}&=&\int_0^\infty \frac{\mu^2\, d\mu}{\cos \left( \frac{2 \pi k}{n} \right) + \cosh (\pi \mu )}
  = \frac{2 i\, \big[ \operatorname{Li}_3\!\big( - e^{\frac{2 i k \pi}{n}} \big) 
  - \operatorname{Li}_3\!\big( - e^{- \frac{2 i k \pi}{n}} \big) \big]}{\pi^3 \sin\!\big(\frac{2k\pi}{n}\big)} 
  \\[1.5mm] 
  \nonumber
  &=&-\frac{i\, \log\big(e^{\frac{2 i k \pi}{n}} \big) \big[ \pi^2+\log^2\big(e^{\frac{2 i k \pi}{n}} \big) \big]}{3\pi^3 \sin\!\big(\frac{2k\pi}{n}\big)} 
  =\frac{2}{\sin\!\big(\frac{2k\pi}{n}\big)} \times\left\{ 
  \begin{array}{ll}
\frac{k(n^2-4k^2)}{3n^3}\,, & k < n / 2 \\[1mm]
-\frac{(n-k)(n-2k)(3n-2k)}{3n^3}\,, & k > n / 2 \,.
\end{array} \right. 
\eea
Above,  we manipulated the tri-logarithm $\operatorname{Li}_3$ using the polylog identity 
\be
\operatorname{Li}_3(z) - \operatorname{Li}_3\!\big(z^{-1}\big) = - \frac{1}{6} \log^3 \left( - z \right) - \frac{\pi^2}{6} 
\log \left( - z \right) \,,
\label{Trilogarithms}
\ee
which holds for $z \notin \,]0,1[$.

Combining the results \reef{I1} and \reef{I2}, we find that the result is the same for $1<k<n/2$ and $n/2<k<n$, namely
\be
\int_0^\infty d\mu\ \frac{(n - 2 k)^2 + \mu^2 n^2}{\cos \left( \frac{2 \pi k}{n} \right) + \cosh (\pi \mu )}
~=~(n - 2 k)^2  I^{(1)}_{n;k} + n^2 I^{(2)}_{n;k}
~=~\frac{8k(n-k)(n-2k)}{3n \,\sin\!\big(\frac{2k\pi}{n}\big)}\,.
\ee
Thus, having evaluated the integral in \reef{sigmaBnow}, we can write $\sigma_n^{(B)}$ as the finite sum
\be\label{resRenyiB}
\sigma_n^{(B)} = \frac{1}{24 \pi\ n^3 \left( n - 1 \right)} \sum_{k = 1}^{n - 1} k \left( n-k  \right) \left( n- 2 k  \right) \tan\! \left( \tfrac{\pi k}{n} \right) \,.
\ee
Note that taking the limit $k \to n/2$ as described below \reef{sigman-res}, the summand evaluates precisely to the special case \reef{speck}. The expression \reef{resRenyiB} is the result for the R\'enyi corner coefficient   presented in \reef{sigman-res}, so this completes our evaluation for the free scalar.

\vspace{2mm}
\noindent {\bf Free fermion.} 
For the fermion field, the R\'{e}nyi corner  coefficient is  given by the integral
\be
\sigma_n^{(F)} = - \frac{2}{n - 1} \sum_{k>0}^{\frac{1}{2} (n - 1)} \int_0^\infty d\mu\ \left[ a \left( 1 - a \right) \mu^2 H - \frac{\mu F}{4 \pi} \right]_{a = k / n}\,,
\ee
where  the sum is over $k$ from 
$1/2$ ($n$ even) or  $1$ ($n$ odd) in integer steps 
to $\frac{1}{2} (n - 1)$.
Substituting the expressions for $H$ and $F$ obtained earlier gives 
\begin{multline}
\sigma_n^{(F)} = \sum_{k>0}^{\frac{1}{2} (n - 1)} \frac{\sin^2 \left( \frac{\pi k}{n} \right)}{8 \pi (n - 1)} 
\int_0^\infty d\mu\ \frac{2 \mu^2 \left[ \cos \left( \frac{2 \pi k}{n} \right) + \cosh (\pi \mu ) \right] - \mu \left( \frac{4 k^2}{n^2} + \mu^2 - 1 \right) \pi \sinh (\pi \mu)}{\left[ \cos \left( \frac{2 \pi k}{n} \right) + \cosh (\pi \mu ) \right]^2}\,.
\end{multline}
We then use integration by parts to simplify the integral
\begin{multline}
\sigma_n^{(F)} = \sum_{k>0}^{\frac{1}{2} (n - 1)}
\frac{\sin^2 \left( \frac{\pi k}{n} \right)}{8 \pi (n - 1)} 
\int_0^\infty d\mu\ \left[ \frac{\partial}{\partial \mu} \left( \frac{\mu \left( \frac{4 k^2}{n^2} + \mu^2 - 1 \right)}{\cos \left( \frac{2 \pi k}{n} \right) + \cosh (\pi \mu )} \right) - \frac{\frac{4 k^2}{n^2} + \mu^2 - 1}{\cos \left( \frac{2 \pi k}{n} \right) + \cosh (\pi \mu )} \right].
\end{multline}
The boundary term integrates to zero and the expression simplifies to
\be
\sigma_n^{(F)} = \sum_{k>0}^{\frac{1}{2} (n - 1)} \frac{\sin^2 \left( \frac{\pi k}{n} \right)}{8 \pi (n - 1)} \int_0^\infty d\mu\ \frac{1 - \mu^2 - \frac{4 k^2}{n^2}}{\cos \left( \frac{2 \pi k}{n} \right) + \cosh (\pi \mu )}\,.
\ee
The result of the integral again involves a difference of two tri-logarithms and it can be simplified using equation~(\ref{Trilogarithms}). The result is even in $k \to -k$ and we can write the final answer as
\be
\sigma_n^{(F)} = \frac{1}{24 \pi\ n^3 \left( n - 1 \right)}\sum_{k = - \frac{1}{2} (n - 1)}^{\frac{1}{2} (n - 1)} k \left( n^2 - 4 k^2 \right) \tan \left( \frac{\pi k}{n} \right).
\ee
This is the answer we presented in \reef{sigman-resF}. Values for low $n$ were tabulated in section \ref{sec:Introduction} for both $\sigma_n^{(B)}$ and $\sigma_n^{(F)}$.

\section{Asymptotic behavior of the R\'enyi's}
\label{sec:AB} 
Let us now study the large $n$ behavior of the R\'enyi entropy corner coefficients $\sigma_n$. In particular, we evaluate analytically the value for the coefficients $\sigma_n$ in the limit where $n \to \infty$. This is done by introducing a new variable $x = k / n$ and multiplying by $n \Delta x = 1$. Then in the $n \to \infty$ limit, the sum becomes an integral and we have
\be
 \begin{split}
\sigma_\infty^{(B)} &=~ \frac{1}{24 \pi} \int_0^1 dx\ x \left( x - 1 \right) \left( 2 x - 1 \right) \tan \left( \pi x \right) = \frac{3 \zeta(3)}{32 \pi^4}\,,\\[2mm]
\sigma_\infty^{(F)} &=~ \frac{1}{24 \pi} \int_{-1/2}^{1/2} dx\ x \left( 1 - 4 x^2 \right) \tan \left( \pi x \right) = \frac{\zeta(3)}{4 \pi^4}\,.
\end{split}
\ee
These values turn out to be proportional to the asymptotic values of the $\mathcal{F}_n \to n \mathcal{F}_\infty$ calculated on the $n$-covered 3-sphere \cite{Klebanov:2011uf}; as noted in \reef{sigmaF} we have
$\sigma_\infty^{(B/F)} = \frac{1}{4\pi^2}
\mathcal{F}_\infty^{(B/F)}$.

On the right in figure \ref{sigmaplot}, we illustrated the asymptotic behavior of the corner coefficient which we find to be
\be
  \label{largenexp}
  \sigma_n =\sigma_\infty \Big(1  + \frac{b_1}{n}  + \frac{b_2}{n^2} + \frac{b_3}{n^3}+ \dots\Big) \,.
\ee
Numerical fits show that $b_1$, $b_2$, and $b_3$ are $1$ for the free boson while $b_1$ is $1$ and $b_2= b_3\approx 0.31578$ in \reef{largenexp} for the free fermion. In fact, fitting up to  $O(1/n^{16})$, we find numerical evidence that $b_{2k}=b_{2k+1}$ for both the scalar and fermion.  This indicates that a factor of $(n+1)/n$ can be factored out of the function in \reef{largenexp}, so that 
\be
  \label{largenexp2}
  \sigma_n =\sigma_\infty \frac{n+1}{n}\Big(1  + \frac{b_2}{n^2}  + \frac{b_4}{n^4} + \frac{b_6}{n^6}+ \dots\Big) \,.
\ee

It is also interesting to study the ratios of the R\'enyi corner coefficients at large $n$: based on numerical fits in the range $n=100$ to $2000$ we find
\be
\frac{\sigma_n^{(B)}}{\sigma_n^{(F)}} = \frac{3}{8} \Big[ 1 
+ \frac{\pi^2}{12 \zeta(3)} \frac{1}{n^2} 
- 0.93871149 \frac{1}{n^4} 
+ O\Big(\frac{1}{n^5}\Big) \Big]\,.
\ee 
The value of the $1/n^2$-coefficient is guessed based on the numerics. Specifically, we fit to the function 
\be
  \label{fitfunc}
   \frac{3}{8} \Big(1  + \frac{d_1}{n}  + \frac{d_2}{n^2} + \frac{d_3}{n^3}+ \dots\Big)\,,
\ee
and find that $d_1 < 10^{-26}$, $\big|d_2-\frac{\pi^2}{12 \zeta(3)}\big| < 10^{-23}$, $d_3 < 10^{-19}$, $d_4 = -0.93871149\dots$, $d_5 < 10^{-13}$ etc. The vanishing of the odd powers  in \reef{fitfunc} is consistent with \reef{largenexp2}. Note also that we can now identify the number $b_2= b_3\approx 0.31578$ from the fit \reef{largenexp} of the free fermion R\'enyi entropy corner coefficient at large $n$ as $1-\tfrac{\pi^2}{12 \zeta(3)}$; this is the value given in the caption of figure \ref{sigmaplot}.

Taking the Hurwitz zeta-function expressions for $\mathcal{F}_n^{(B/F)}$ from \cite{Klebanov:2011uf} and using \reef{fitfunc} to perform a similar fit at large $n$  in the range $30$ to $300$, we find
\be
\frac{\mathcal{F}_n^{(B)}}{\mathcal{F}_n^{(F)}} = \frac{3}{8} \Big[ 1 
- \frac{\pi^2}{12 \zeta(3)} \frac{1}{n^2}  
+ 0.937106586 \frac{1}{n^4} 
+ O\Big(\frac{1}{n^5}\Big) \Big] \,.
\ee 
Again, the value of the $1/n^2$-coefficient is guessed based on the numerics which give $d_1 < 10^{-20}$, $\big|d_2+\frac{\pi^2}{12 \zeta(3)}\big| < 10^{-17}$, $d_3 < 10^{-14}$, $d_4 = 0.937106586\dots$, $d_5 < 10^{-12}$ etc. The behaviors of $\mathcal{F}_n^{(B/F)}$ individually is, however, very different that that of the R\'enyi corner coefficients. We find that $\mathcal{F}_n^{(B)} \sim n \mathcal{F}_\infty^{(B)} \big( 1+ O(\tfrac{1}{n^4})\big)$ while 
$\mathcal{F}_n^{(F)} \sim n \mathcal{F}_\infty^{(F)} \big( 1+ \tfrac{\pi^2}{12 \zeta(3)} \tfrac{1}{n^2} +O(\tfrac{1}{n^4})\big)$.

It is not clear whether the similarities observed at large $n$ between $\sigma_n$ and $\mathcal{F}_n$  have any significance or if it is a coincidence. Perhaps future investigations will clarify this.

\section*{Acknowledgments}

We are grateful to Horacio Casini for sharing with us the integral expression for the corner coefficient $\sigma_n$ of the free fermion. 
We would like to thank Pablo Bueno, Finn Larsen, Rob Myers, Silviu Pufu, and William Witczak-Krempa  for useful discussions. HE is supported in part by NSF CAREER Grant PHY-0953232 and she is a Cottrell Scholar of the Research Corporation for Science Advancement. MH is supported by a Fulbright Fellowship and by the Department of Physics at the University of Michigan.




\begin{thebibliography}{99}

\bibitem{Fradkin:2006mb} 
  E.~Fradkin and J.~E.~Moore,
  ``Entanglement entropy of 2D conformal quantum critical points: hearing the shape of a quantum drum,''
  Phys.\ Rev.\ Lett.\  {\bf 97}, 050404 (2006)
  [cond-mat/0605683 [cond-mat.str-el]].

\bibitem{Casini:2006hu} 
  H.~Casini and M.~Huerta,
  ``Universal terms for the entanglement entropy in 2+1 dimensions,''
  Nucl.\ Phys.\ B {\bf 764}, 183 (2007)
  [hep-th/0606256].
    
\bibitem{Hirata:2006jx} 
  T.~Hirata and T.~Takayanagi,
  ``AdS/CFT and strong subadditivity of entanglement entropy,''
  JHEP {\bf 0702}, 042 (2007)
  [hep-th/0608213].

\bibitem{Casini:2008as} 
  H.~Casini, M.~Huerta and L.~Leitao,
  ``Entanglement entropy for a Dirac fermion in three dimensions: Vertex contribution,''
  Nucl.\ Phys.\ B {\bf 814}, 594 (2009)
  [arXiv:0811.1968 [hep-th]].
  
\bibitem{Casini:2009sr} 
  H.~Casini and M.~Huerta,
  ``Entanglement entropy in free quantum field theory,''
  J.\ Phys.\ A {\bf 42}, 504007 (2009)
  [arXiv:0905.2562 [hep-th]].
  
  
  

\bibitem{Bueno:2015rda} 
  P.~Bueno, R.~C.~Myers and W.~Witczak-Krempa,
  ``Universality of corner entanglement in conformal field theories,''
  arXiv:1505.04804 [hep-th].



\bibitem{Bueno:2015xda} 
  P.~Bueno and R.~C.~Myers,
  ``Corner contributions to holographic entanglement entropy,''
  arXiv:1505.07842 [hep-th].


\bibitem{Osborn:1993cr} 
  H.~Osborn and A.~C.~Petkou,
  ``Implications of conformal invariance in field theories for general dimensions,''
  Annals Phys.\  {\bf 231}, 311 (1994)
  [hep-th/9307010].
  
\bibitem{Klebanov:2011uf} 
  I.~R.~Klebanov, S.~S.~Pufu, S.~Sachdev and B.~R.~Safdi,
  ``Renyi Entropies for Free Field Theories,''
  JHEP {\bf 1204}, 074 (2012)
  [arXiv:1111.6290 [hep-th]].
  
  
\end{thebibliography}
\end{document}